
\magnification \magstep1
\raggedbottom
\openup 4\jot
\voffset6truemm
\headline={\ifnum\pageno=1\hfill\else
\hfill {\it One-Loop Amplitudes in Euclidean Quantum Gravity}
\hfill \fi}
\def\cstok#1{\leavevmode\thinspace\hbox{\vrule\vtop{\vbox{\hrule\kern1pt
\hbox{\vphantom{\tt/}\thinspace{\tt#1}\thinspace}}
\kern1pt\hrule}\vrule}\thinspace}
\centerline {\bf ONE-LOOP AMPLITUDES IN EUCLIDEAN QUANTUM GRAVITY}
\vskip 0.1cm
\centerline {Giampiero Esposito,$^{1,2}$ Alexander Yu.
Kamenshchik,$^{3}$}
\centerline {Igor V. Mishakov,$^{3}$ and Giuseppe Pollifrone$^{4}$}
\vskip 0.1cm
\centerline {\it ${ }^{1}$Istituto Nazionale di Fisica Nucleare,
Sezione di Napoli,}
\centerline {\it Mostra d'Oltremare Padiglione 20, 80125 Napoli, Italy}
\centerline {\it ${ }^{2}$Dipartimento di Scienze Fisiche,
Mostra d'Oltremare Padiglione 19, 80125 Napoli, Italy}
\centerline {\it ${ }^{3}$Nuclear Safety Institute, Russian Academy
of Sciences,}
\centerline {\it 52 Bolshaya Tulskaya, Moscow, 113191, Russia}
\centerline {\it ${ }^{4}$Dipartimento di Fisica, Universit\`a
di Roma ``La Sapienza", and}
\centerline {\it INFN, Sezione di Roma, Piazzale Aldo Moro 2,
00185 Roma, Italy}
\vskip 0.1cm
\noindent
This paper studies the linearized
gravitational field in the presence of boundaries. For this
purpose, $\zeta$-function regularization is used to perform the
mode-by-mode evaluation of BRST-invariant Faddeev-Popov
amplitudes in the case of flat Euclidean four-space
bounded by a three-sphere. On choosing the de Donder
gauge-averaging term, the resulting $\zeta(0)$ value is found
to agree with the space-time covariant calculation of the same
amplitudes, which relies on the recently corrected geometric
formulas for the asymptotic heat kernel in the case of mixed
boundary conditions. Two sets of mixed boundary conditions
for Euclidean quantum gravity are then compared in detail.
The analysis proves that one cannot restrict the
path-integral measure to transverse-traceless perturbations.
By contrast, gauge-invariant amplitudes are only obtained
on considering from the beginning all perturbative modes
of the gravitational field, jointly with ghost modes.
\vskip 0.1cm
\leftline {PACS number(s): 03.70.+k, 04.60.Ds, 98.80.Hw}
\vskip 100cm
\centerline{\bf I. INTRODUCTION}
\vskip 1cm
In the Euclidean functional integral approach to quantum
gravity [1], one deals with amplitudes written formally
as path integrals over all Riemannian four-geometries
matching the boundary data on (compact) Riemannian
three-geometries $\Bigr(\Sigma_{1},h_{1}\Bigr)$ and
$\Bigr(\Sigma_{2},h_{2}\Bigr)$. To take into account the
gauge freedom of the theory, the path-integral measure
also includes suitable ghost fields, described geometrically
by a one-form, hereafter denoted by $\varphi=\varphi_{\mu}
dx^{\mu}$ (see appendix), subject to boundary conditions at
$\Bigr(\Sigma_{1},h_{1}\Bigr)$ and $\Bigr(\Sigma_{2},h_{2}\Bigr)$.
Although a rigorous definition of the Feynman sum [2] over all
Riemannian four-geometries with their topologies does not
yet exist, the choice of boundary conditions still plays a key
role to obtain a well-defined elliptic boundary-value
problem, which may be applied to the semiclassical analysis of
the quantum theory.

In quantum cosmology, it was proposed in Refs. [3,4] that no
boundary conditions should be imposed at the three-geometry
$\Bigr(\Sigma_{1},h_{1}\Bigr)$, since this might shrink to a
point in the case of the quantum state of the universe. One
would then have to impose suitable boundary conditions only
at $\Bigr(\Sigma_{2},h_{2}\Bigr)$, by describing the quantum
state of the universe in terms of a Euclidean path integral
over all {\it compact} Riemannian four-geometries matching
the boundary data at $\Bigr(\Sigma_{2},h_{2}\Bigr)$. Although
this approach to quantum cosmology still involves a number
of formal definitions, the semiclassical evaluation of the
corresponding wave function may be put on solid grounds.
The one-loop analysis
is related to mathematical and physical subjects such
as cobordism theory (i.e. under which conditions a compact
manifold is the boundary of another compact manifold),
the geometry of compact Riemannian
four-manifolds, the asymptotic heat kernel, the one-loop effective
action and the use of mixed boundary conditions in
quantum field theory (see below).

In particular, over the last ten years many efforts have been
produced to evaluate one-loop
quantum amplitudes for gauge fields and the gravitational
field in the presence of boundaries, either by using
the space-time covariant
Schwinger-DeWitt method [5] or the mode-by-mode
analysis wich relies on $\zeta$-function
regularization [6,7].

The main motivations were the
need to understand the relation between different approaches to
quantum field theories in the
presence of boundaries, and the quantization of closed cosmologies.
Indeed, boundaries play an important role in the Feynman path-integral
approach to quantum gravity [8] as we just said, in choosing
BRST-covariant and  gauge-invariant boundary conditions
for quantum cosmology [9,10]
and in studying different quantizazion
and regularization techniques in field theory.
In particular, for the latter problem,
discrepancies were found in the semiclassical evaluation of
quantum amplitudes by using space-time covariant methods, where
the scaling factor of one-loop quantum amplitudes
coincides with the Schwinger-DeWitt
$A_{2}$ coefficient in the heat-kernel
expansion, or noncovariant methods, where the same
factor can be computed
within the framework of $\zeta$-function regularization, and it is
expressed through the $\zeta(0)$ value.

If one reduces a field theory with first-class constraints to its
physical degrees of freedom before quantization [11-14],
one of the main problems is whether
the resulting quantum theory is equivalent to the
theories relying on the Faddeev-Popov gauge-averaging method or
on the extended-phase-space Hamiltonian path integral of Batalin,
Fradkin and Vilkovisky, where one takes into account ghost and
gauge modes [13-14]. We will see that, in a manifestly
gauge-invariant formulation of such theories, there seem to be
no unphysical modes, in that there are no subsets of the
set of all perturbative modes whose effects cancel exactly
the ones of ghost modes.
This lack of cancellation turns out to be essential to
achieve agreement between different techniques (see Sec. IV).

In Ref. [11],
the $\zeta(0)$ calculation was performed
for gravitons by restricting the path-integral measure
to transverse-traceless perturbations
in the case of flat Euclidean four-space
bounded by a three-sphere. In Refs. [15-17], this result was
generalized to the part of the Riemannian de Sitter
four-sphere bounded by a three-sphere. Both results did
not coincide with those obtained by a space-time covariant
method [9]. Hence the natural hypothesis arises that the
possible non-cancellation of the contributions of gauge
and ghost modes can be the cause of the discrepancy.
In Refs. [18-21] such a suggestion was checked for the
electromagnetic field on different manifolds and in
different gauges.

In Ref. [22] the asymptotic heat kernel for second-order
elliptic operators was obtained in the case
of pure and mixed boundary conditions in real Riemannian
four-manifolds, and in Ref. [23] this analysis has been
improved. In the light of these results,
the conformal anomalies on
Einstein spaces with boundaries
have been re-calculated in Ref. [24].

In Ref. [25], we calculated the $\zeta(0)$ value for
gravitons in the de Donder gauge
on the part of flat four-dimensional
Euclidean space bounded by two concentric three-spheres,
taking into account the contribution of gauge modes
and ghosts. The result was in agreement with the
space-time covariant calculation.
Hence we here investigate
the linearized gravitational field
in the geometric framework of Ref. [11]
(i.e. flat Euclidean space bounded by a three-sphere),
and we compare the resulting $\zeta(0)$ value
with the space-time covariant calculation of the same
Faddeev-Popov
amplitudes, by using the recently corrected geometric formulas
for the asymptotic heat kernel in the case of mixed
boundary conditions [24].
For our purposes, we use the version of the $\zeta$-function
tecnique [6] elaborated in Refs. [15-17].
Hence we write
$f_{n}(M^{2})$ for the function occurring in the equation obeyed by
the eigenvalues by virtue of boundary conditions, and $d(n)$ for the
degeneracy of the eigenvalues parametrized by the integer $n$. One then
defines the function
$$
I(M^{2},s) \equiv \sum_{n=n_{0}}^{\infty} d(n) \; n^{-2s}
\; {\rm ln} \; f_{n}(M^{2}) \; .
\eqno (1.1)
$$
Such a function has an analytic continuation to the whole complex-$s$
plane as a meromorphic function, i.e.
$$
``I(M^{2},s)"={I_{\rm pole}(M^{2})\over s}+I^{R}(M^{2})
+O(s) \; .
\eqno (1.2)
$$
Then the $\zeta(0)$ value is
$$
\zeta(0)=I_{\rm log}+I_{\rm pole}(\infty)-I_{\rm pole}(0)
\; ,
\eqno (1.3)
$$
where $I_{\rm log}$ is the coefficient of $\log \; M$ from
$I(M^{2},s)$ as $M \rightarrow \infty$, and $I_{\rm pole}(M^{2})$
is the residue at $s=0$. The uniform asymptotic
expansions of basis functions as both their order and $M$
tend to $\infty$, yield $I_{\rm log}$ and $I_{\rm pole}(\infty)$,
while the limiting form of basis functions as $M \to 0$ and
$n \to \infty$ yields $I_{\rm pole}(0)$ [15-17,19-21,25].

In Sec. II we consider the mixed boundary conditions
of Refs. [9,10] and compute
the $\zeta(0)$ value for the linearized
gravitational field on the part of flat four-dimensional
Euclidean space bounded by a three-sphere, taking into
account the contributions of gauge modes and ghosts.
Sec. III studies instead the mixed boundary conditions
first proposed in Ref. [26]. Sec. IV compares
the results of Sec. II
with the ones deriving from a geometric analysis of the
asymptotic heat kernel.
Concluding remarks are presented in Sec. V, and relevant
details are given in the appendix.
\vskip 1cm
\centerline {\bf II. LUCKOCK-MOSS-POLETTI}
\centerline {\bf BOUNDARY CONDITIONS}
\vskip 1cm
In this section we
evaluate $\zeta(0)$ for
the linearized gravitational field in the de Donder gauge [25]
$$
\Phi_{\nu}^{dD}(h) \equiv \nabla^{\mu}
\Bigr(h_{\mu \nu}-{1\over 2}g_{\mu \nu}g^{\rho \sigma}
h_{\rho \sigma}\Bigr) \; ,
\eqno (2.1)
$$
where $h$ is the perturbation of the background
four-metric $g$, and $\nabla$ is the Levi-Civita
connection compatible with $g$, i.e. $\nabla g =0$.
This gauge-averaging functional leads to the familiar
form of the elliptic operator acting on metric
perturbations. For example, in the case of a flat Euclidean
background, this operator reduces to
$-\cstok{\ } \equiv -\nabla^{\mu}\nabla_{\mu}$ [25].
Our background four-geometry is indeed flat
Euclidean space bounded by a three-sphere,
studied also in  Ref. [11], and we take
into account the contributions of gauge modes and ghosts.

In Refs. [9,10] mixed boundary conditions have been
introduced for quantum gravity and quantum cosmology.
They are motivated by the need to obtain
BRST-covariant boundary conditions which lead to
gauge-invariant quantum amplitudes (see, however, Sec. V).
This request leads to Dirichlet boundary conditions
on the normal component $\varphi_{0}$ of the ghost
one-form, to Robin conditions
on the spatial components $\varphi_{i}$
of the ghost one-form
and to mixed boundary conditions on the
perturbative modes
of the gravitational field.

The Luckock-Moss-Poletti (also referred to as LMP)
boundary conditions read ($n$ being the normal to the
boundary) [9,10,25]
$$
\Bigr[h_{ij}\Bigr]_{\partial M}=0 \; ,
\eqno (2.2)
$$
$$
\Bigr[h_{0i}\Bigr]_{\partial M}=0 \; ,
\eqno (2.3)
$$
$$
\left[\left(2K_{\sigma}^{\sigma}+
n^{\sigma}\nabla_{\sigma}\right)n^{\mu}n^{\nu}
\Bigr(h_{\mu\nu}-{1 \over 2}g_{\mu\nu}g^{\alpha\beta}
h_{\alpha\beta}\Bigr)\right]_{\partial M}=0 \; ,
\eqno (2.4)
$$
$$
\Bigr[\varphi_{0}\Bigr]_{\partial M}=0 \; ,
\eqno (2.5)
$$
$$
\Bigr[\left(- K_{\mu}^{\; \; \nu}
+ \delta_{\mu}^{\; \; \nu} \; n^{\rho}
\nabla_{\rho}\right)
P_{\nu}^{\sigma}\varphi_{\sigma}\Bigr]_{\partial M}=0 \; ,
\eqno (2.6)
$$
where $K_{\mu\nu}$ and $P_{\mu}^{\nu}\equiv \delta_{\mu}^{\nu} -
n_{\mu}n^{\nu}$ are the extrinsic-curvature
tensor of the boundary and the tangential
projection operator respectively.

The mode-by-mode form of the LMP boundary conditions
appears already in Ref. [25], and hence we here re-write them
only for the spatial components $\varphi_{i}$ of the ghost, i.e.
$$
\biggr[{\partial\varphi_{i} \over \partial\tau} -
{2 \over \tau}\varphi_{i} \biggr]_{\partial M}=0 \; ,
\eqno (2.7)
$$
to correct an unfortunate mistake in Ref. [25], where
the numerical coefficient of $1 \over \tau$ is incorrectly given
(cf. Eqs. (3.3), (3.12)-(3.13) therein).
That mistake did not affect the $\zeta(0)$ value for ghosts,
but it would affect the one-boundary calculations we are studying.
On inserting the mode-by-mode expansion
of the components $\varphi_{i}(x,\tau)$
of the ghost one-form
(see the appendix and Ref. [25])
into the boundary conditions (2.7) one obtains
$$
{dm_{n}\over d\tau}-{2\over \tau}m_{n}=0
\; \; \; \; {\rm at} \; \; \; \; {\partial M} \; ,
\eqno (2.8)
$$
$$
{dp_{n}\over d\tau}-{2\over \tau}p_{n}=0
\; \; \; \; {\rm at} \; \; \; \; {\partial M} \; .
\eqno (2.9)
$$
Following the notation of the appendix and
the technique used in Ref. [25],
one finds the following contributions
to $\zeta(0)$:
$$
\zeta(0)_{\rm transverse-traceless \; modes}=-{278\over 45} \; ,
\eqno (2.10)
$$
$$
\zeta(0)_{\rm scalar \; modes}=18-{1\over 60}
+{5\over 2} -9-{1\over 6}-{1\over 180}
={509 \over 45} \; ,
\eqno (2.11)
$$
$$
\zeta(0)_{\rm partially \; decoupled \; modes}=-2-17=-19 \; ,
\eqno (2.12)
$$
$$
\zeta(0)_{\rm vector \; modes}=15-{41\over 60}
-{2\over 3}-{31\over 180}={1213\over 90} \; ,
\eqno (2.13)
$$
$$
\zeta(0)_{\rm decoupled \; vector \; mode}=-{21\over 2} \; ,
\eqno (2.14)
$$
$$
\zeta(0)_{\rm scalar \; ghost \; modes}=-2\biggr({119\over 120}
+{899\over 360}\biggr)=-{314\over 45} \; ,
\eqno (2.15)
$$
$$
\zeta(0)_{\rm vector \; ghost \; modes}=-2\biggr({19\over 120}
+{209\over 360}\biggr)=-{133\over 90} \; ,
\eqno (2.16)
$$
$$
\zeta(0)_{\rm decoupled \; ghost \; mode}={5\over 2} \; .
\eqno (2.17)
$$
For example, in (2.11) we evaluate
$I_{\rm log}$ by using the uniform asymptotic
expansions of modified Bessel fuctions.
After eliminating fake roots of order $(4n-1)$ as
$M \to 0$ one finds that the coefficient of $\ln M$ is
$-(4n+1)$. Hence one obtains
$$
I_{\rm log}=\sum_{n=3}^{\infty}{n^{2}\over 2}(-1-4n)
=-2\zeta_{R}(-3)+18+{5 \over 2}=-{1\over 60}+18+{5 \over 2} \; .
\eqno (2.18)
$$

To evaluate $I_{\rm pole}(\infty)$,
the structure of the resulting $n$-dependent coefficient
in the eigenvalue condition is
$$
\sigma_{\infty}(n)=12n {(n^{2}-1)\over (n^{2}-4)} \; ,
\eqno (2.19)
$$
and it is easy to see that the term
${n^{2}\over 2}{\rm ln}(\sigma_{\infty}(n))$
does not contribute as $n \to \infty$ (i.e. it has no
coefficient of ${1 \over n}$),
and hence $I_{\rm pole}(\infty)=0$.

Last, as $M \rightarrow 0$ and $n \rightarrow \infty$,
the $n$-dependent coefficient which contributes to
$I_{\rm pole}(0)$ is
$$
\sigma_{0}(n)=12\Gamma^{-4}(n){{(n+1)(n+4)(n-1)^{2}}
\over {n^{3}(n-2)(n+2)^{2}}}  \; .
\eqno (2.20)
$$
Thus, using Stirling's asymptotic
expansion of the $\Gamma$-function [27],
and after taking the coefficient of $1 \over n$
as $n \to \infty$ in
${n^{2}\over 2}{\rm ln}(\sigma_{0}(n))$,
one finds
$$
I_{\rm pole}(0)=9+{1 \over 6} + {1 \over 180} \; .
\eqno (2.21)
$$

Finally, by virtue of (2.10)-(2.17) one gets the full
$\zeta(0)$ value for gravitons
$$
\zeta(0)=-{758\over 45} \; .
\eqno (2.22)
$$
Hence the contributions of gauge and ghost modes do not cancel
each other (cf. Ref. [11]) and
our result, as we will show in Sec. IV, coincides with that
obtained by using the covariant Schwinger-DeWitt
technique on the part of flat Euclidean four-space bounded by
a three-sphere.
\vskip 1cm
\centerline {\bf III. BARVINSKY BOUNDARY CONDITIONS}
\vskip 1cm
The boundary conditions studied in Sec. II are not the only
possible set of mixed boundary conditions for quantum
gravity. As shown in Ref. [26], one can also set to zero
at the boundary the gauge-averaging functional, the whole
ghost one-form, and the perturbation of the induced three-metric.
With the notation of Sec. II, after making a {\it gauge}
transformation of the metric perturbation $h_{\mu \nu}$
according to the law [11]
$$
{\hat h}_{\mu \nu} \equiv h_{\mu \nu}
+\nabla_{(\mu} \; \varphi_{\nu)} \; ,
\eqno (3.1)
$$
one finds in the de Donder gauge
(denoting by $\lambda$ the eigenvalues of the
elliptic operator $-g_{\mu \nu}\cstok{\ }-R_{\mu \nu}$)
$$
\Phi_{\nu}^{dD}(h)-\Phi_{\nu}^{dD}(\hat h)=-{1\over 2}
\Bigr(g_{\mu \nu}\cstok{\ } +R_{\mu \nu}\Bigr)\varphi^{\mu}
={\lambda \over 2}\varphi_{\nu} \; ,
\eqno (3.2)
$$
for any background with Ricci tensor $R_{\mu \nu}$. In
our flat Euclidean background, the Ricci tensor vanishes,
and on making a 3+1 split of the de Donder functional
$\Phi_{\nu}^{dD}$ and of the ghost one-form
$\varphi_{\mu}$, the boundary conditions proposed in
Ref. [26] read
$$
\Bigr[h_{ij}\Bigr]_{\partial M}
=\Bigr[{\hat h}_{ij}\Bigr]_{\partial M}=0 \; ,
\eqno (3.3)
$$
$$
\Bigr[\Phi_{0}^{dD}(h)\Bigr]_{\partial M}
=\Bigr[\Phi_{0}^{dD}(\hat h)\Bigr]_{\partial M}=0 \; ,
\eqno (3.4)
$$
$$
\Bigr[\Phi_{i}^{dD}(h)\Bigr]_{\partial M}
=\Bigr[\Phi_{i}^{dD}(\hat h)\Bigr]_{\partial M}=0 \; ,
\eqno (3.5)
$$
$$
\Bigr[\varphi_{0}\Bigr]_{\partial M}=0 \; ,
\eqno (3.6)
$$
$$
\Bigr[\varphi_{i}\Bigr]_{\partial M}=0 \; .
\eqno (3.7)
$$
Note that the vanishing of the whole ghost one-form at the
boundary ensures the invariance of the boundary
conditions (3.3) under the transformations
(3.1) (see also Eq. (5.1)). At that
stage, the only remaining set of boundary conditions on
metric perturbations, whose invariance under (3.1) is again
guaranteed by (3.6)-(3.7), is given by (3.4)-(3.5)
by virtue of (3.2).
In this respect, these boundary conditions are the
natural generalization of magnetic boundary conditions for
Euclidean Maxwell theory, where one sets to zero at the
boundary the tangential components of the potential, the
gauge-averaging functional, and hence the ghost zero-form
[18-21]. The boundary conditions (3.3)-(3.7) were considered
in Ref. [26] as part of the effort to understand the relation
between the wave function of the universe and the effective
action in quantum field theory. The loop expansion in
quantum cosmology was then obtained after a thorough study
of boundary conditions for the propagator [26].

In the light of (3.3), the boundary conditions (3.4)-(3.5)
lead to mixed boundary conditions on the metric perturbations
which take the form
$$
\left[{\partial h_{00}\over \partial \tau}
+{6\over \tau}h_{00}-{\partial \over \partial \tau}
\Bigr(g^{ij}h_{ij}\Bigr)+{2\over \tau^{2}}
h_{0i}^{\; \; \; \mid i} \right]_{\partial M}=0 \; ,
\eqno (3.8)
$$
$$
\left[{\partial h_{0i}\over \partial \tau}
+{3\over \tau}h_{0i}-{1\over 2}
{\partial h_{00}\over \partial x^{i}}\right]_{\partial M}
=0 \; .
\eqno (3.9)
$$
To evaluate the scaling behaviour of the corresponding
one-loop amplitudes, it is necessary to write down the
mode-by-mode form of the boundary conditions (3.8)-(3.9),
(3.3) and (3.6)-(3.7). They lead to (see Eqs.
(A1)-(A13) of the appendix)
$$
{da_{n}\over d\tau}+{6\over \tau}a_{n}-{1\over \tau^{2}}
{de_{n}\over d\tau}-{2\over \tau^{2}}b_{n}=0
\; \; \; \; {\rm at} \; \; \; \; {\partial M} \; ,
\eqno (3.10)
$$
$$
{db_{n}\over d\tau}+{3\over \tau}b_{n}
-{(n^{2}-1)\over 2}a_{n}=0 \; \; \; \; {\rm at} \; \; \; \;
{\partial M} \; ,
\eqno (3.11)
$$
$$
{dc_{n}\over d\tau}+{3\over \tau}c_{n}=0
\; \; \; \; {\rm at} \; \; \; \; {\partial M} \; ,
\eqno (3.12)
$$
$$
d_{n}=0 \; \; \; \; {\rm at} \; \; \; \; {\partial M} \; ,
\eqno (3.13)
$$
$$
e_{n}=0 \; \; \; \; {\rm at} \; \; \; \; {\partial M} \; ,
\eqno (3.14)
$$
$$
f_{n}=0 \; \; \; \; {\rm at} \; \; \; \; {\partial M} \; ,
\eqno (3.15)
$$
$$
k_{n}=0 \; \; \; \; {\rm at} \; \; \; \; {\partial M} \; ,
\eqno (3.16)
$$
$$
l_{n}=0 \; \; \; \; {\rm at} \; \; \; \; {\partial M} \; ,
\eqno (3.17)
$$
$$
m_{n}=0 \; \; \; \; {\rm at} \; \; \; \; {\partial M} \; ,
\eqno (3.18)
$$
$$
p_{n}=0 \; \; \; \; {\rm at} \; \; \; \; {\partial M} \; .
\eqno (3.19)
$$
On using the technique outlined in the introduction
and applied also in Sec. II and in Ref. [25], the corresponding
contributions to $\zeta(0)$ are found to be (see the appendix
for details)
$$
\zeta(0)_{\rm transverse-traceless \; modes}=-{278\over 45} \; ,
\eqno (3.20)
$$
$$
\zeta(0)_{\rm scalar \; modes}=18-{1\over 60}-1-{1\over 180}
={764\over 45} \; ,
\eqno (3.21)
$$
$$
\zeta(0)_{\rm partially \; decoupled \; modes}=-2-15=-17 \; ,
\eqno (3.22)
$$
$$
\zeta(0)_{\rm vector \; modes}=12-{11\over 60}
-{2\over 3}-{31\over 180}={494\over 45} \; ,
\eqno (3.23)
$$
$$
\zeta(0)_{\rm decoupled \; vector \; mode}=-{15\over 2} \; ,
\eqno (3.24)
$$
$$
\zeta(0)_{\rm scalar \; ghost \; modes}=-2\biggr({179\over 120}
+{59\over 360}\biggr)=-{149\over 45} \; ,
\eqno (3.25)
$$
$$
\zeta(0)_{\rm vector \; ghost \; modes}=-2\biggr(-{41\over 120}
-{31\over 360}\biggr)={77\over 90} \; ,
\eqno (3.26)
$$
$$
\zeta(0)_{\rm decoupled \; ghost \; mode}={5\over 2} \; .
\eqno (3.27)
$$
Hence the full $\zeta(0)$ for linearized gravity subject
to the Barvinsky boundary conditions [26] is found to be
$$
\zeta(0)=-{241\over 90} \; .
\eqno (3.28)
$$
Note that the result (3.28) differs from the one obtained
in the previous section: $\zeta(0)=-{758\over 45}$. As far
as we can see, this property reflects two different
physical situations described by two different sets of
mixed boundary conditions (see also the comments in Sec. V).
In the following section, the
result (2.22) is checked by using geometric
formulas for the asymptotic heat kernel.
\vskip 10cm
\centerline {\bf IV. GEOMETRIC RESULTS ON THE HEAT KERNEL}
\vskip 1cm
In this section we calculate by
means of a geometric method the
Schwinger-DeWitt coefficient $A_{2}$
for the case of Luckock-Moss-Poletti
boundary conditions. The result obtained coincides with
the one found in Sec. II by a mode-by-mode analysis.
The space-time covariant formulas for a
second-order elliptic operator on a manifold
with boundaries in the case of
pure and mixed boundary conditions were obtained
in Ref. [22] and corrected in Ref. [23]. We use them in the
form presented in Ref. [24].

The Schwinger-DeWitt coefficient $A_{2}$
for the elliptic operator
$$
-D_{\mu}D^{\mu} + X \; ,
$$
where $D_{\mu}$ is a gauge derivative with curvature $F_{\mu \nu}$,
can be written as
$$
16 \pi^{2} A_{2} = \int_{\cal M} b_{2} d\mu +\int_{\partial {\cal M}}
c_{2} d\mu \; .
\eqno (4.1)
$$
The volume coefficient $b_{2}$ is well-known [5], while surface terms
depend upon the choice of boundary conditions. We use mixtures
of Dirichlet and Robin boundary conditions,
$$
P_{-}\phi = 0 \; ,\ \ \ (\psi
+ n^{\sigma} \nabla_{\sigma}) P_{+} \phi = 0 \; ,
\eqno (4.2)
$$
where $P_{\pm}$ are projection operators [9,10,24].

The results can be expressed in terms of polynomials in the curvature
tensor $R_{\mu\nu\alpha\beta}$ of the background four-manifold and
in terms of the extrinsic-curvature tensor of the boundary
(hereafter $R$ is the trace of the Ricci tensor, and $K$ is
the trace of $K_{\mu \nu}$), i.e.
$$ \eqalignno{
q& \equiv
{8 \over 3} K^{3} + {16 \over 3} K_{\mu}^{\nu} K_{\nu}^{\alpha}
K_{\alpha}^{\mu}
- 8 K K_{\mu\nu} K^{\mu\nu} + 4 K R \cr
&- 8 R_{\mu\nu}
(K n^{\mu} n^{\nu} + K^{\mu\nu}) + 8 R_{\mu\nu\alpha\beta}
K^{\mu\alpha} n^{\nu} n^{\beta} \; ,
&(4.3)\cr}
$$
and
$$
p \equiv K_{\mu}^{\nu} K_{\nu}^{\alpha} K_{\alpha}^{\mu} -
K K_{\mu\nu} K^{\mu\nu} + {2 \over 9} K^{3} \; .
\eqno (4.4)
$$
For Dirichlet boundary conditions [24]
$$
c_{2}^{D} = {\rm Tr} \biggr[-{1\over 360} q + {2\over 35} p -
{1\over 3} \Bigr(X - {1\over 6} R \Bigr) K
- {1\over 2} n^{\sigma} \nabla_{\sigma}
\Bigr(X- {1 \over 6} R \Bigr)
+ {1\over 15} C_{\mu\nu\alpha\beta} K^{\mu\alpha}
n^{\nu} n^{\beta} \biggr] \; ,
\eqno (4.5)
$$
while for Robin boundary conditions [24]
$$ \eqalignno{
c_{2}^{R}&= {\rm Tr} \biggr[-{1\over 360} q + {2\over 45} p -
{1\over 3} \Bigr(X - {1\over 6} R \Bigr) K
+ {1 \over 2} n^{\sigma} \nabla_{\sigma}
\Bigr(X- {1\over 6} R \Bigr)
- {4\over 3} \Bigr(\psi - {1\over 3}K \Bigr)^{3}\cr
&+2 \Bigr(X - {1\over 6}R \Bigr) \psi
+ \Bigr(\psi - {1\over 3}K \Bigr)
\Bigr({2\over 45} K^{2}
- {2\over 15} K_{\mu\nu} K^{\mu\nu}\Bigr)
+ {1\over 15} C_{\mu\nu\alpha\beta} K^{\mu\alpha}
n^{\nu} n^{\beta} \biggr] \; .
&(4.6)\cr}
$$
For mixed boundary conditions [23,24]
$$ \eqalignno{
c_{2}&= {\rm Tr} \Bigr[P_{+} c_{2}^{R}
+ P_{-} c_{2}^{D} -{2\over 15}
(P_{+|i})(P_{+}^{|i}) K
- {4\over 15} (P_{+|i})(P_{+|j})K^{ij}\cr
&+ {4\over 3} (P_{+|i})(P_{+}^{|i}) P_{+} \psi
- {2\over 3} P_{+}
(P_{+}^{|i}) n ^{\mu} F_{i\mu}\Bigr] \; ,
&(4.7)\cr}
$$
where Latin indices run from 1 to 3, Greek indices run
from 0 to 3, and the stroke denotes (as in the rest of
our paper) covariant differentiation tangentially with
respect to the three-dimensional Levi-Civita connection
of the boundary [22]. The Luckock-Moss-Poletti
boundary conditions for gravitons imply that [9]
$$
\Bigr(P_{+ {\rm gravitons}}\Bigr)_{\mu \nu}^{\alpha \beta}
\equiv n_{\mu} n_{\nu} \Bigr(2 n^{\alpha} n^{\beta} -
g^{\alpha\beta}\Bigr) \; ,
\eqno (4.8)
$$
and
$$
\psi_{\rm gravitons} = 2 K \; .
\eqno (4.9)
$$
Similarly, for ghosts one has
$$
\Bigr(P_{+ {\rm ghosts}}\Bigr)_{\mu}^{\nu}
\equiv \delta_{\mu}^{\nu} - n_{\mu} n^{\nu} \; ,
\eqno (4.10)
$$
and
$$
\psi_{\rm ghosts} = - {K \over 3} \; .
\eqno (4.11)
$$
Note that, on taking traces for gravitons in Eq. (4.7),
one has to use the generalized Kronecker symbol
$$
\delta_{\mu\nu}^{\alpha \beta} \equiv
{1\over 2} \biggr(\delta_{\mu}^{\alpha} \; \delta_{\nu}^{\beta}
+\delta_{\mu}^{\beta} \; \delta_{\nu}^{\alpha} \biggr) \; .
$$
Now inserting (4.8) and (4.9) into Eqs. (4.1), (4.6)-(4.7)
one obtains for gravitons
$$
A_{2_{\rm gravitons}} = - {98 \over 9} \; .
\eqno (4.12)
$$
Analogously, using the expressions (4.10)-(4.11) one finds for ghosts
$$
A_{2_{\rm ghosts}} = -{268 \over 45} \; .
\eqno (4.13)
$$
These results are in full agreement with those in Sec. II.
\vskip 1cm
\centerline {\bf V. CONCLUSIONS}
\vskip 1cm
Our paper has studied one-loop quantum gravity with boundaries
in the limiting case of small three-geometries, which is
relevant for quantum cosmology [11]. This leads to the analysis
of flat Euclidean four-space bounded by a three-sphere,
with mixed boundary conditions on metric perturbations. The
results of our investigation are as follows.

First, we have completed and improved the analysis of
Luckock-Moss-Poletti mixed boundary conditions [9,10]
considered in our earlier work [25]. These are motivated by
the analysis of BRST transformations at the boundary,
and are more relevant for supersymmetric theories of
gravitation [7,10]. The eight contributions to the full
$\zeta(0)$ for Faddeev-Popov amplitudes have been calculated
in detail by means of a mode-by-mode analysis, and the
resulting $\zeta(0)$ value has been found to agree with the
geometric theory of the asymptotic heat kernel. The latter
relies on the results appearing in Refs. [22-24].

Second, we have performed a detailed analysis of the mixed
boundary conditions for quantum gravity proposed by
Barvinsky in Ref. [26]. Their main merit is the invariance
under the transformations (3.1) on metric perturbations.
The technique of $\zeta$-function
regularization, jointly with a mode-by-mode analysis, has
made it possible to obtain the full $\zeta(0)$ value as in
Eq. (3.28). Indeed, this result differs from the one obtained
in Sec. II. Thus, different boundary conditions lead to
different semiclassical wavefunctions, with the exception of
Euclidean Maxwell theory, where magnetic or electric boundary
conditions correspond to the same semiclassical amplitudes
in the Lorentz gauge [20-21]. It should be emphasized that,
in Sec. II, the invariance of the boundary conditions (2.3)
under (3.1) leads to the Robin boundary conditions (2.6)
on the ghost one-form. However, this implies that the
boundary conditions (2.2) are {\it not} invariant under (3.1),
since
$$
{\hat h}_{ij}-h_{ij}=\varphi_{(i \mid j)}
+K_{ij}\varphi_{0} \; .
\eqno (5.1)
$$
In other words, the right-hand side of (5.1) does not vanish at
$\partial M$ if (2.6) holds, because it reduces to
$\Bigr[\varphi_{(i \mid j)}\Bigr]_{\partial M}$ by virtue of
(2.5). Indeed, in Ref. [9] the authors acknowledge that the
boundary conditions (2.2)-(2.6) are not entirely BRST
invariant, but they say, without explicit proof, that this
does not affect the gauge invariance of the resulting quantum
amplitudes. Hence we found it appropriate to consider also
the boundary conditions of Sec. III, which are instead
completely invariant under the transformations (3.1).
On the other hand, the boundary conditions of Sec. II are
motivated by self-adjointness theory [9] and are in
agreement with the results on boundary conditions for
one-forms, which should be mixed for gauge fields [23,28].
Hence the boundary conditions of Sec. III are less natural
in this respect.

Third, our analysis proves that, also in the one-boundary
problem, one cannot restrict the path-integral measure to
transverse-traceless perturbations. By contrast, ghost modes
and the whole set of perturbative modes of the gravitational
field are all necessary to obtain gauge-invariant
(one-loop) amplitudes.

It now remains to be seen what happens on considering
arbitrary relativistic gauges and curved backgrounds. As
one already knows from Euclidean Maxwell theory, it is not
possible to diagonalize the operator matrix of the problem
for arbitrary gauge conditions [21]. In this respect, it appears
interesting to characterize {\it all} relativistic gauge
conditions which enable one to express the perturbative
modes as linear combinations of Bessel functions [20,21,25].
Moreover, it is necessary to obtain geometric
formulas for $\zeta(0)$ in the case of Barvinsky boundary
conditions studied in Sec. III. These boundary conditions
are not naturally expressed in terms of projection operators,
and hence the method of Refs. [22-24] cannot be applied.
Nevertheless, work is in progress on this problem by the
author of Refs. [23,28] and his collaborators. Last, but not
least, the analysis of non-relativistic gauges for
Euclidean quantum gravity within the Faddeev-Popov
formalism, and the non-local nature of the one-loop effective
action, deserve careful consideration.

If one wants to relate our analysis to the Lorentzian
theory, there is also the problem of interpreting the quantum
state corresponding to the boundary conditions of Secs.
II and III. Moreover, the impossibility to restrict the
measure of the Euclidean path integral to transverse-traceless
perturbations raises further interpretive issues for the
Lorentzian theory, where such a reduction to physical degrees
of freedom is instead quite natural. Thus, although the
path-integral formulation of Euclidean quantum gravity [29]
does not provide a mathematically consistent theory of the
quantized gravitational field, the detailed calculations and
the open problems presented in our paper seem to add evidence
in favour of quantum cosmology having a deep influence on
modern quantum field theory in four-dimensions [7,30].
\vskip 10cm
\centerline {\bf ACKNOWLEDGMENTS}
\vskip 1cm
We are indebted to Andrei Barvinsky for suggesting to study
his boundary conditions, and to Dmitri Vassilevich for
correspondence. Our research was partially supported
by the European Union under the Human Capital and Mobility
Program. Moreover, the research described in this publication
was made possible in part by Grant No MAE000 from the
International Science Foundation. The work of A.Yu. Kamenshchik
was partially supported by the Russian Foundation for
Fundamental Researches through grant No 94-02-03850-a,
and by the Russian Research Project ``Cosmomicrophysics".
\vskip 1cm
\centerline {\bf APPENDIX}
\vskip 1cm
Following Ref. [25], the metric perturbations are expanded
on a family of three-spheres centred on the origin as
$$
h_{00}(x,\tau)=\sum_{n=1}^{\infty}a_{n}(\tau)Q^{(n)}(x) \; ,
\eqno (A1)
$$
$$
h_{0i}(x,\tau)=\sum_{n=2}^{\infty}
\left[b_{n}(\tau){Q_{\mid i}^{(n)}(x)\over (n^{2}-1)}
+c_{n}(\tau)S_{i}^{(n)}(x)\right] \; ,
\eqno (A2)
$$
$$ \eqalignno{
h_{ij}(x,\tau)&=\sum_{n=3}^{\infty}d_{n}(\tau)
\left({Q_{\mid ij}^{(n)}(x)\over (n^{2}-1)}
+{c_{ij}\over 3}Q^{(n)}(x)\right)\cr
&+\sum_{n=1}^{\infty}{e_{n}(\tau)\over 3}c_{ij}Q^{(n)}(x)\cr
&+\sum_{n=3}^{\infty}\biggr[f_{n}(\tau)
\Bigr[S_{i\mid j}^{(n)}(x)+S_{j \mid i}^{(n)}(x)\Bigr] \cr
&+k_{n}(\tau)G_{ij}^{(n)}(x)\biggr] \; ,
&(A3)\cr}
$$
where $Q^{(n)}(x),S_{i}^{(n)}(x),G_{ij}^{(n)}(x)$ are
scalar, transverse vector, and transverse-traceless
tensor hyperspherical harmonics, respectively, on a
unit three-sphere with metric $c_{ij}$. The components
$\varphi_{0}(x,\tau)$ and $\varphi_{i}(x,\tau)$ of the
ghost one-form are expanded as in (A1)-(A2), providing one
replaces $a_{n}(\tau),b_{n}(\tau),c_{n}(\tau)$ by
the modes $l_{n}(\tau),m_{n}(\tau)$ and $p_{n}(\tau)$
respectively [25]. Note that, strictly, our ghost one-form
corresponds to a ghost $\eta_{\mu}$ and an anti-ghost
${\overline \eta}_{\mu}$. They obey the same boundary
conditions introduced for $\varphi_{\mu}$ [10], and their
contribution to $\zeta(0)$ is obtained by using the
multiplicative factor -2, as in Secs. II-IV.

In the one-boundary problems studied in our paper, regular
perturbative modes are only obtained by setting to zero the
coefficients multiplying the modified Bessel functions
$K_{n}$, since such functions are singular at
the origin of flat Euclidean four-space.
Hence one finds for scalar-type
gravitational perturbations (cf. Ref. [25])
$$
a_{n}(\tau) = {1\over\tau}\Bigr[\gamma_{1}I_{n}(M\tau)
+\gamma_{3}I_{n-2}(M\tau) + \gamma_{4}I_{n+2}(M\tau)\Bigr] \; ,
\eqno (A4)
$$
$$
b_{n}(\tau) =\gamma_{2}I_{n}(M\tau) + (n + 1)
\gamma_{3}I_{n-2}(M\tau) - (n-1)\gamma_{4}I_{n+2}(M\tau) \; ,
\eqno (A5)
$$
$$
d_{n}(\tau) =\tau \left[-\gamma_{2}I_{n}(M\tau) +
{(n+1)\over(n-2)}\gamma_{3}I_{n-2}(M\tau) +
{(n-1)\over(n+2)}\gamma_{4}I_{n+2}(M\tau)\right] \; ,
\eqno (A6)
$$
$$
e_{n}(\tau) =\tau
\Bigr[3\gamma_{1}I_{n}(M\tau) - 2\gamma_{2}I_{n}(M\tau)
-\gamma_{3}I_{n-2}(M\tau) - \gamma_{4}I_{n+2}(M\tau)
\Bigr] \; .
\eqno (A7)
$$
The basis functions for vectorlike gravitational
perturbations are
$$
c_{n}(\tau) = {\widetilde \varepsilon}_{1}I_{n+1}(M\tau) +
{\widetilde \varepsilon}_{2}I_{n-1}(M\tau) \; ,
\eqno (A8)
$$
$$
f_{n}(\tau) = \tau\left[-{1 \over (n+2)}
{\widetilde \varepsilon}_{1}I_{n+1}(M\tau)
+ {1\over (n-2)}{\widetilde \varepsilon}_{2}
I_{n-1}(M\tau)\right] \; ,
\eqno (A9)
$$
and the basis function for transverse-traceless
symmetric tensor harmonics reads
$$
k_{n}={\alpha}_{1} \tau I_{n}(M\tau)  \; .
\eqno (A10)
$$
Finally the basis functions for ghosts are
$$
l_{n}(\tau) = {1\over \tau}\Bigr[\kappa_{1}I_{n+1}(M\tau) +
\kappa_{2}I_{n-1}(M\tau)\Bigr] \; ,
\eqno (A11)
$$
$$
m_{n}(\tau) = - (n-1)\kappa_{1}I_{n+1}(M\tau) +
(n+1)\kappa_{2}I_{n-1}(M\tau) \; ,
\eqno (A12)
$$
$$
p_{n}(\tau) = \vartheta I_{n}(M\tau) \; .
\eqno (A13)
$$
The equations (A4)-(A13) hold both in Sec. II and in Sec. III,
and we here focus on the latter application, since the former
was treated in detail in Ref. [25].

On inserting the form of $a_{n}(\tau),b_{n}(\tau),
d_{n}(\tau),e_{n}(\tau)$ into the boundary conditions
(3.10)-(3.11) and (3.13)-(3.14), one gets an eigenvalue
condition for coupled scalar modes given by the vanishing of
the determinant of a $4 \times 4$ matrix. As
$M \rightarrow 0$, fake roots of order $4n$ are found to
arise. In the calculation of $I_{\rm log}$, the factors
$\sqrt{M}$ and ${1\over \sqrt{M}}$ compensate each other as
$M \rightarrow \infty$, and hence $I_{\rm log}$ is found
to be
$$
I_{\rm log}=\sum_{n=3}^{\infty}{n^{2}\over 2}(-4n)
=-2\zeta_{R}(-3)+18=-{1\over 60}+18 \; .
\eqno (A14)
$$
Moreover, as $M \rightarrow \infty$ and $n \rightarrow \infty$,
the $n$-dependent coefficient in the eigenvalue condition
takes the form (cf. Eq. (2.19))
$$
\rho_{\infty}(n)=12n {(n^{2}-1)\over (n^{2}-4)} \; .
\eqno (A15)
$$
Since ${\rho_{\infty}(n)\over n}$ is an even function of $n$, the
term ${n^{2}\over 2}{\rm ln}(\rho_{\infty}(n))$ has no
coefficient of ${1\over n}$ as $n \rightarrow \infty$,
and hence $I_{\rm pole}(\infty)=0$.

By contrast, as $M \rightarrow 0$ and $n \rightarrow \infty$,
the $n$-dependent coefficient in the eigenvalue condition
takes the form (cf. Eq. (2.20))
$$
\rho_{0}(n)=\Gamma^{-4}(n)\biggr(1-{1\over n}\biggr)
{48\over (n+1)(n-2)} \; .
\eqno (A16)
$$
Thus, by virtue of Stirling's asymptotic expansion [27]
$$
{\rm ln} \; \Gamma(n) \sim
\biggr(n-{1\over 2}\biggr){\rm ln}(n)
-n+{1\over 2}{\rm ln}(2\pi)+{1\over 12}{1\over n}
-{1\over 360}{1\over n^{3}}+{\rm O}(n^{-5}) \; ,
\eqno (A17)
$$
the $I_{\rm pole}(0)$ value, which is the coefficient of
${1\over n}$ in the asymptotic expansion as
$n \rightarrow \infty$ of ${n^{2}\over 2}{\rm ln}(\rho_{0}(n))$,
turns out to be
$$
I_{\rm pole}(0)={1\over 180}-{1\over 6}+{4\over 3}-{1\over 6}
={181\over 180} \; .
\eqno (A18)
$$

The same technique yields the contributions (3.22)-(3.27)
to the full $\zeta(0)$, bearing in mind that, for decoupled
modes, the contribution to $\zeta(0)$ is only given by the
$I_{\rm log}$ coefficient [15-17].
\vskip 10cm
\hrule
\vskip 1cm

\item {[1]}
S.W. Hawking, in {\it General Relativity, an Einstein
Centenary Survey}, edited by S.W. Hawking and W. Israel
(Cambridge University Press, Cambridge, 1979).

\item {[2]}
R.P. Feynman, Acta Physica Polonica {\bf 24}, 697 (1963).

\item {[3]}
J.B. Hartle and S.W. Hawking, Phys. Rev. D {\bf 28},
2960 (1983).

\item {[4]}
S.W. Hawking, Nucl. Phys. {\bf B239}, 257 (1984).

\item {[5]}
B.S. DeWitt, {\it Dynamical Theory of Group and Fields}
(Gordon and Breach, New York, 1965).

\item {[6]}
S.W. Hawking, Commun. Math. Phys. {\bf 55}, 133 (1977).

\item {[7]}
G. Esposito, {\it Quantum Gravity, Quantum Cosmology and
Lorentzian Geometries}, Lecture Notes in Physics: Monographs,
Vol. m12 (Springer-Verlag, Berlin, 1994).

\item {[8]}
C.W. Misner, Rev. Mod. Phys. {\bf 29}, 497 (1957).

\item {[9]}
I.G. Moss and S.J. Poletti, Nucl. Phys. {\bf B341}, 155 (1990).

\item {[10]}
H.C. Luckock, J. Math. Phys. {\bf 32}, 1755 (1991).

\item {[11]}
K. Schleich, Phys. Rev. D {\bf 32}, 1889 (1985).

\item {[12]}
J. Louko, Phys. Rev. D {\bf 38}, 478 (1988).

\item {[13]}
M. Henneaux, Phys. Rep. {\bf 126}, 1 (1985);
D. McMullan and I. Tsutsui, Ann. Phys. (N.Y.)
{\bf 237}, 269 (1995).

\item {[14]}
M. Henneaux and C. Teitelboim, {\it Quantization of Gauge Systems}
(Princeton University Press, Princeton, 1992).

\item {[15]}
A.O. Barvinsky, A.Yu. Kamenshchik, and I.P. Karmazin,
Ann. Phys. (N.Y.) {\bf 219}, 201 (1992).

\item {[16]}
A.O. Barvinsky,
A.Yu. Kamenshchik, I.P. Karmazin, and I.V. Mishakov,
Class. Quantum Grav. {\bf 9}, L27 (1992).

\item {[17]}
A.Yu. Kamenshchik and I.V. Mishakov,
Int. J. Mod. Phys. A {\bf 7}, 3713 (1992).

\item {[18]}
G. Esposito, Class. Quantum Grav. {\bf 11}, 905 (1994).

\item {[19]}
G. Esposito and A.Yu. Kamenshchik, Phys. Lett. B
{\bf 336}, 324 (1994).

\item {[20]}
G. Esposito, A.Yu. Kamenshchik, I.V. Mishakov, and
G. Pollifrone, Class. Quantum Grav {\bf 11}, 2939 (1994).

\item {[21]}
G. Esposito, A.Yu. Kamenshchik, I.V. Mishakov, and
G. Pollifrone, ``Relativistic Gauge Conditions in Quantum
Cosmology'' DSF Report No. 95/8 (unpublished).

\item {[22]}
T.P. Branson and P.B. Gilkey, Commun. Part. Diff. Eq. {\bf 15},
245 (1990).

\item {[23]}
D.V. Vassilevich, ``Vector Fields on a Disk with Mixed
Boundary Conditions'' St. Petersburg Report No. SPbU-IP-94-6
(unpublished).

\item {[24]}
I.G. Moss and S.J. Poletti, Phys. Lett. B {\bf 333},
326 (1994).

\item {[25]}
G. Esposito, A.Yu. Kamenshchik, I.V. Mishakov, and
G. Pollifrone, Phys. Rev. D {\bf 50}, 6329 (1994).

\item {[26]}
A.O. Barvinsky, Phys. Lett. B {\bf 195}, 344 (1987).

\item {[27]}
{\it Handbook of Mathematical Functions with Formulas,
Graphs and Mathematical Tables}, edited by
M. Abramowitz and I. Stegun, Natl. Bur. Stand. Appl.
Math. Ser. No. 55 (U.S. GPO, Washington, D.C., 1965).

\item {[28]}
D.V. Vassilevich, ``QED on Curved Background and on
Manifolds with Boundaries: Unitarity versus Covariance''
ICTP Report No. 94/359 (unpublished).

\item {[29]}
{\it Euclidean Quantum Gravity}, edited by G.W. Gibbons
and S.W. Hawking (World Scientific, Singapore, 1993).

\item {[30]}
M. Bordag, E. Elizalde, and K. Kirsten, ``Heat-Kernel
Coefficients of the Laplace Operator on the
D-dimensional Ball'' UB-ECM-PF Report No. 95/3 (unpublished).

\bye